\documentclass{physeauth}
\usepackage{graphicx}
\usepackage{amsmath}
\usepackage{amssymb}

\newcommand{\eq}[1]{Eq.~(\ref{#1})}

\newcommand{\bref}[1]{Ref.~\cite{#1}}

\begin{document}

\begin{frontmatter}

\title{Application of new R\'enyi uncertainty relations to
wave packet revivals} 

\author[efm,ic1]{Francisco de los Santos \thanksref{thank1}},
\author[efm]{Clara Guglieri} and
\author[famn,ic1]{Elvira Romera}

\address[efm]{Departamento de Electromagnetismo y F{\'\i}sica de la
Materia, Universidad de Granada, Fuentenueva s/n, 18071 Granada,
Spain}
\address[famn]{Departamento de F\'isica At\'omica, Molecular y Nuclear, 
Universidad de Granada, Fuentenueva s/n, 18071 Granada, Spain}
\address[ic1]{Instituto Carlos I de F{\'\i}sica Te\'orica y
Computacional, Universidad de Granada, Fuentenueva s/n, 18071 Granada,
Spain}

\thanks[thank1]{
Corresponding author. 
E-mail: dlsantos@onsager.ugr.es}

\begin{abstract}
Wave packet revivals and fractional revivals are studied by means of newly
derived uncertainty relations that involve R\'enyi entropies and position and momentum 
dispersions.

\end{abstract}

\begin{keyword}
Quantum information \sep wave packet dynamics \sep dephasing and revivals 
\PACS 42.50.Md, 03.67.-a, 3.65.Ge
\end{keyword}
\end{frontmatter}


\section{Introduction}

The time evolution of quantum wave packets may lead to interesting collapse and revival phenomena. 
Propagating wave packets initially evolve quasiclassically and oscillate
with a classical period $T_{\rm cl}$, but eventually spread and collapse. At later times, multiples of the `revival time' $T_{\rm rev}$, wave
packets regain their initial wave form and behave quasiclassically again.
Additionally, at times that are rational fractions of $T_{\rm rev}$, the
wave packet temporarily splits into a number of scaled copies called 
fractional revivals \cite{3,4,rob}. Revivals and fractional revivals have attracted a great interest during the past decades. They have been investigated theoretically in nonlinear quantum systems, atoms and molecules \cite{1}, and observed experimentally in, among others, Rydberg atoms, molecular vibrational states or Bose-Einstein condensates \cite{2}. 
Recently, methods for isotope separation \cite{isotope}, number factorization \cite{primenumbers} as well as for wave packet control \cite{wp_control} have been put forward that are based on revival phenomena.

It can be shown \cite{3,parker} that the classical period and the revival time of wave
packet evolution are given by the first coefficients of the
Taylor series of the energy spectrum $E_n$ around the energy
$E_{n_0}$ corresponding to the peak of the initial wave packet,
\begin{equation}
E_n \approx E_{n_0}+E'_{n_0} (n-n_0)+\frac{E''_{n_0}}{2} (n-n_0)^2 +\cdots.
\end{equation}
In fact the second-, third- and fourth-order terms in the expansion provide the 
classical period of motion $T_{\rm cl}=2\pi\hbar / |E''_{n_0}|$, the quantum revival scale
time $T_{\rm rev}=4\pi \hbar/|E'_{n_0}|$, and the so-called super-revival time, respectively. 
Fractional revival times are given in terms of the quantum revival scale time \cite{3} by 
$t=p T_{\rm rev}/q $ with $p$ and $q$ mutually prime. 

The study of the time development of wave packet solutions of the
Schr\"odinger equation often makes use of the autocorrelation
function $A(t)$. Within this approach, 
the occurrance of revivals and fractional revivals corresponds to, respectively,
the return of $A(t)$ to its initial value of unity and the appearance of relative maxima in $A(t)$. 
This method, however, misses to detect some fractional revivals because, $A(t)$ being the overlap between the initial wave packet and the evolved one at a given time, the wave packet does not generally regenerate in the same position it started from.
Other methods to study revival phenomena include the time evolution of the expectation values of some quantities \cite{rob,sun,don}, and an approach
based on a finite difference eigenvalue method that allows to predict the revival times directly  \cite{laserphysics}. 

Recently, an information entropy approach has been proposed \cite{rom1}, complementary to the conventional autocorrelation function.
Information entropies measure the spread of the probability density of the wave packet, and therefore can be used with advantage 
to identify the collapse and the regenerating of initially well localized wave packets. Moreover, this approach overcomes the difficulty 
that wave packets reform themselves at locations that do not coincide with their original ones. 
More fully, in terms of the probability densities in position and momentum spaces, $\rho(x)=|\psi(x)|^2$
and $\gamma(p)=|\phi(p)|^2$, respectively, the sum of R\'enyi entropies in conjugate spaces reads
\begin{eqnarray}
R_{\rho}^{(\alpha)} + R_{\gamma}^{(\beta)}= &&
\frac{1}{1-\alpha}\ln \int_{-\infty}^{\infty} \left[\rho(x)\right]^{\alpha} dx \nonumber \\
&+& \frac{1}{1-\beta}\ln \int_{-\infty}^{\infty} \left[\gamma(x)\right]^{\alpha}dx,
\end{eqnarray}
and the R\'enyi uncertainty relation is given by \cite{biarenyi}
\begin{equation}
R_{\rho}^{(\alpha)} + R_{\gamma}^{(\beta)}
\geq -\frac{1}{2(1-\alpha)}\ln\frac{\alpha}{\pi}-\frac{1}{2(1-\beta)}\ln\frac{\beta}{\pi},
\label{uncer}
\end{equation} 
where $1/\alpha+1/\beta=2$. In the limits $\alpha\rightarrow 1$ and $\beta \rightarrow 1$ 
the R\'enyi uncertainty relation (\ref{uncer}) reduces to that of Shannon's \cite{BBM},
$ S_\rho+S_\gamma \geq 1+\ln (\pi)$, which can thus be considered a particular case of the former. 
Within this context, and due to the fact that the uncertainty relation (\ref{uncer}) is saturated 
only for Gaussian wave packets, the temporary formation of fractional revivals 
corresponds to the relative minima of $R_{\rho}^{(\alpha)}(t) + R_{\gamma}^{(\beta)}(t)$.

Although this technique was also shown to be superior to an analysis based on the standard variance uncertainty product \cite{rom2},
\begin{equation}
\sigma_\rho \sigma_\gamma  \geq \hbar/4, 
\end{equation}
with $ \sigma_\rho^2=\langle x^2 \rangle -\langle x \rangle^2$ and $\sigma_\gamma^2=\langle p^2 \rangle -\langle p \rangle^2$,
time dependent expectation values and dispersions provide a more direct connection to
the classical description. It would therefore be of interest if both methods could be combined to
achieve a better description of the phenomenon.
In this paper we show that the analysis of wave packet revivals can be  
carried out using new uncertainty relations involving R\'enyi entropies and momentum and position
dispersions. To be more concrete, we apply the three newly derived relations \cite{nagy,zozor}
\begin{equation}
N^{(\alpha)}_\rho \sigma^2_\gamma \geq D/4, \quad  N^{(\alpha)}_\gamma \sigma^2_\rho \geq D/4, \quad N^{(\alpha)}_\rho N^{(\alpha)}_\gamma \geq 1/4,
\label{uncertainty}
\end{equation}
where $\alpha \in (1/2,1]$, $D$ is the system dimensionality, and $N^{(\alpha)}_f$ is the so-called R\'enyi entropy power of
index $\alpha$, defined as
\begin{equation}
N^{(\alpha)}_f \equiv \Big(\frac{\alpha}{2\alpha-1} \Big)^{\frac{2\alpha -1}{\alpha -1}} \frac{1}{2\pi}e^{2R_f^{(\alpha)}/D}.	 
\end{equation}
Except for the $\alpha=1$ case, the above relations \eq{uncertainty} are not saturated by Gaussians.
Note that the usual R\'enyi uncertainty relation (\ref{uncer}) can be written in terms of the R\'enyi entropy power 
in the compact form 
\begin{equation}
N^{(\alpha)}_\rho N^{(\beta)}_\gamma \geq 1/4.
\end{equation}

Next, we shall consider the uncertainty relations (\ref{uncertainty})
as they apply to revival phenomena, and present the system model we shall be dealing with, 
namely, the so-called quantum `bouncer', that is, a quantum particle bouncing on a hard surface under the
influence of gravity.

\section{Revivals and fractional revivals in the quantum bouncer}

\begin{figure}[tb] 
  \begin{center} 
\includegraphics[angle=0,width=.5\textwidth]{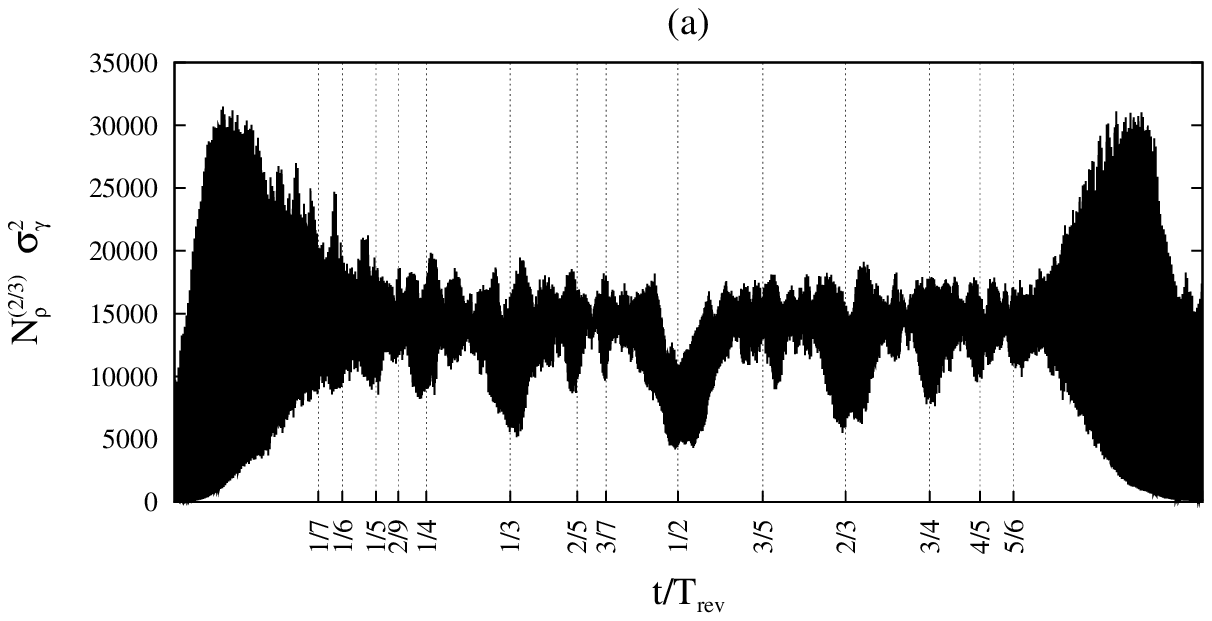}
\includegraphics[angle=0,width=.5\textwidth]{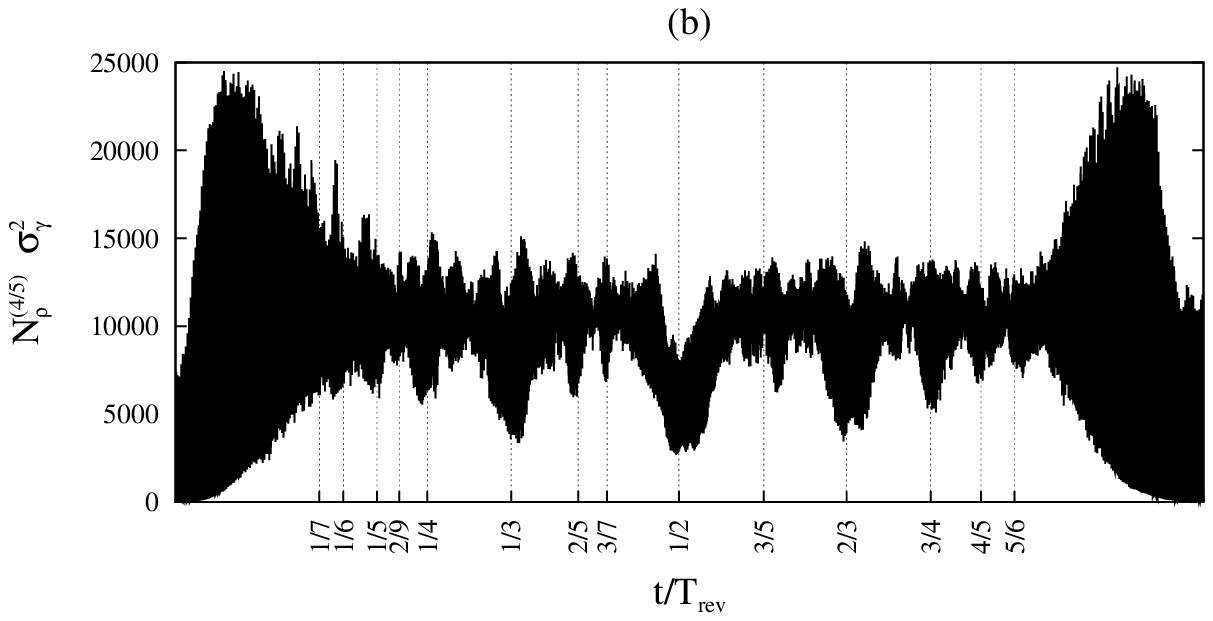}
    \caption{Time dependence of $N^{(\alpha)}_\rho \sigma^2_\gamma$ and main fractional
revivals for an initial Gaussian wave packet with $z_0=100$, $p_0=0$, and $\sigma=1$
in a quantum bouncer. Panel (a) corresponds to $\alpha=2/3$ and (b) to $\alpha=4/5$.}
    \label{fig1} 
  \end{center} 
\end{figure}

Consider an object of mass $m$ that falls towards an impenetrable flat surface
subjected only to the gravitational force directed downward along the $z$ axis,
and described by the potential $V(z)=mgz$, if $z>0$ and  $V(z)=+\infty$ otherwise.
The quantum variant of this familiar classical system has been thoroughly studied
and, interestingly, gravitational quantum bouncers have been recently realized 
using neutrons \cite{qb1} and atomic clouds \cite{qb2}. In the context of wave packet
dynamics, revival behavior has been discussed in \bref{don,gea} and within the
entropic approach in \bref{rom1,rom2}.

\begin{figure}[tb] 
  \begin{center} 
\includegraphics[angle=0,width=.5\textwidth]{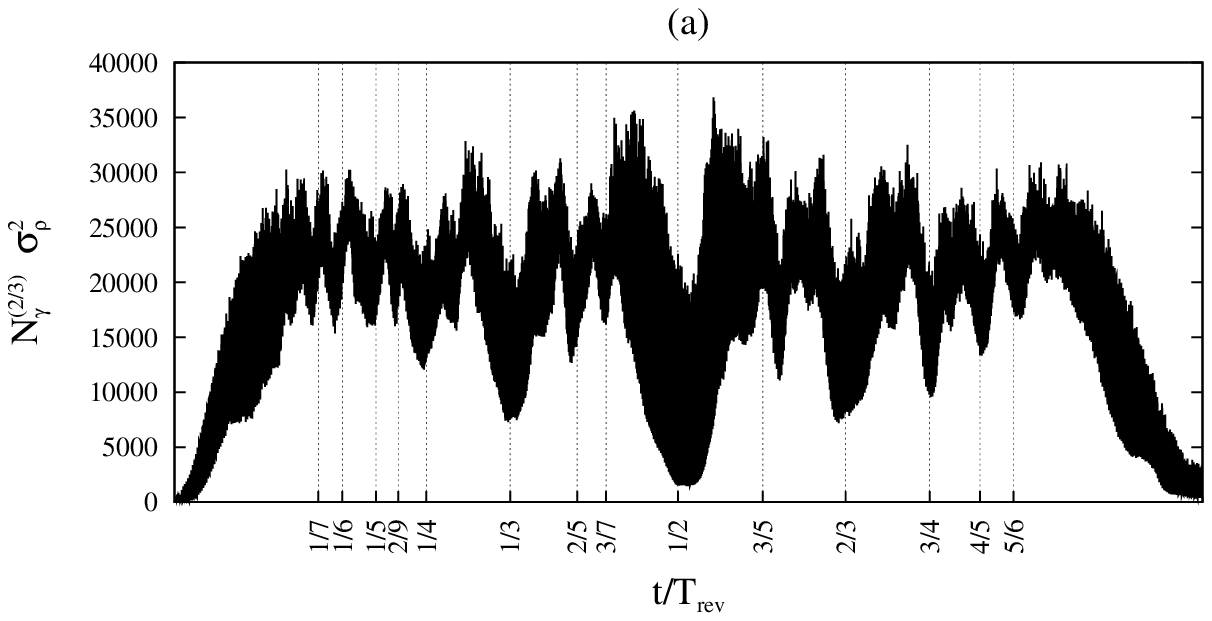}
\includegraphics[angle=0,width=.5\textwidth]{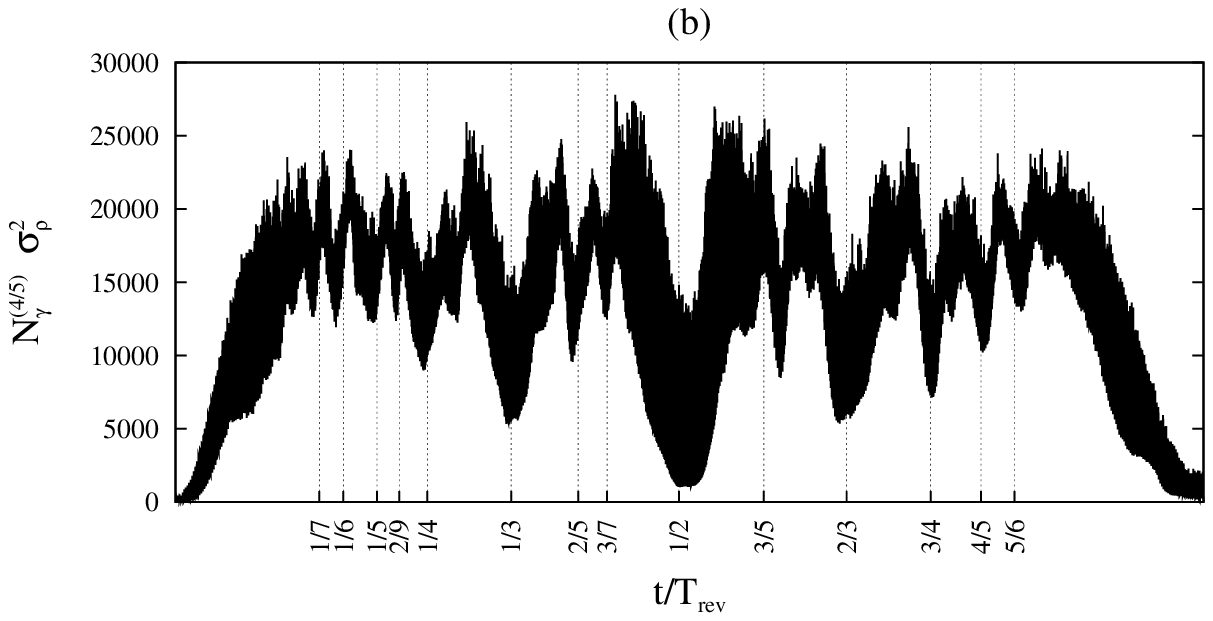}
\caption{Time dependence of $N^{(\alpha)}_\gamma \sigma^2_\rho$ and main fractional
revivals for a quantum bouncer. Parameters as in Fig. \ref{fig1}} 
    \label{fig2} 
  \end{center} 
\end{figure} 

The time dependent solution of the Schr\"odinger equation for the above potential
reads
\begin{equation}
\Psi(z,t) = \sum_{n=1}^\infty C_n e^{-iE_nt/\hbar} \varphi_n(z),
\end{equation}
where the eigenfunctions and eigenvalues are given by \cite{gea}
\begin{equation}
E'_n=z_n; \quad \varphi_n(z')= {\mathcal N}_n
{\rm Ai}(z'-z_n); \quad n=1,2,3,\ldots.
\end{equation} 
Primed symbols denote rescaled position and energy variables $z^{\prime}=z/l_g$, $E^{\prime}=E/mgl_g$,
with $l_g=\left(\hbar/2gm^2\right)^{1/3}$ being a characteristic gravitational length.
${\rm Ai}(z)$ is the Airy function, $-z_n$ denotes its zeros, and ${\mathcal N}_n$ is the $\varphi_n(z^\prime)$ normalization factor. 
In what follows, the primes on the variables are dropped and we assume initial conditions that correspond to Gaussian wave packets
localized at a height $z_0$ above the surface, with a width $\sigma$ and an initial 
momentum $p_0=0$,
\begin{equation}
\Psi(z,0)=\left(\frac{2}{\pi \sigma^2}\right)^{1/4} e^{-(z-z_0)^2/\sigma^2}.
\end{equation}
The corresponding coefficients of the wave function can be obtained analytically as \cite{vallee}, 
\begin{eqnarray}
C_n= &&{\mathcal N}_n \left(2 \pi \sigma ^2\right)^{1/4} 
\exp\left[\frac{\sigma^2}{4}\left(z_0-z_n+\frac{\sigma^4}{24}\right)\right] \nonumber \\
&& \times {\rm Ai}\left(z_0-z_n+\frac{\sigma^4}{16} \right)
\end{eqnarray}
with ${\mathcal N}_n=|{\rm Ai}'(-z_n)|$. Although accurate analytic approximations can be found for
$z_n, C_n$ and ${\mathcal N}_n$ \cite{gea}, these were determined numerically by using scientific 
subroutine libraries for the Airy function. The classical period and the revival time 
can be calculated to obtain $T_{\rm cl}=2\sqrt{z_0}$ and $T_{\rm rev}=4 z_0^2/\pi$, respectively,
and the temporal evolution of the wave packet in momentum-space is obtained numerically by the fast 
Fourier transform method.

\begin{figure}[tb] 
  \begin{center} 
\includegraphics[angle=0,width=.5\textwidth]{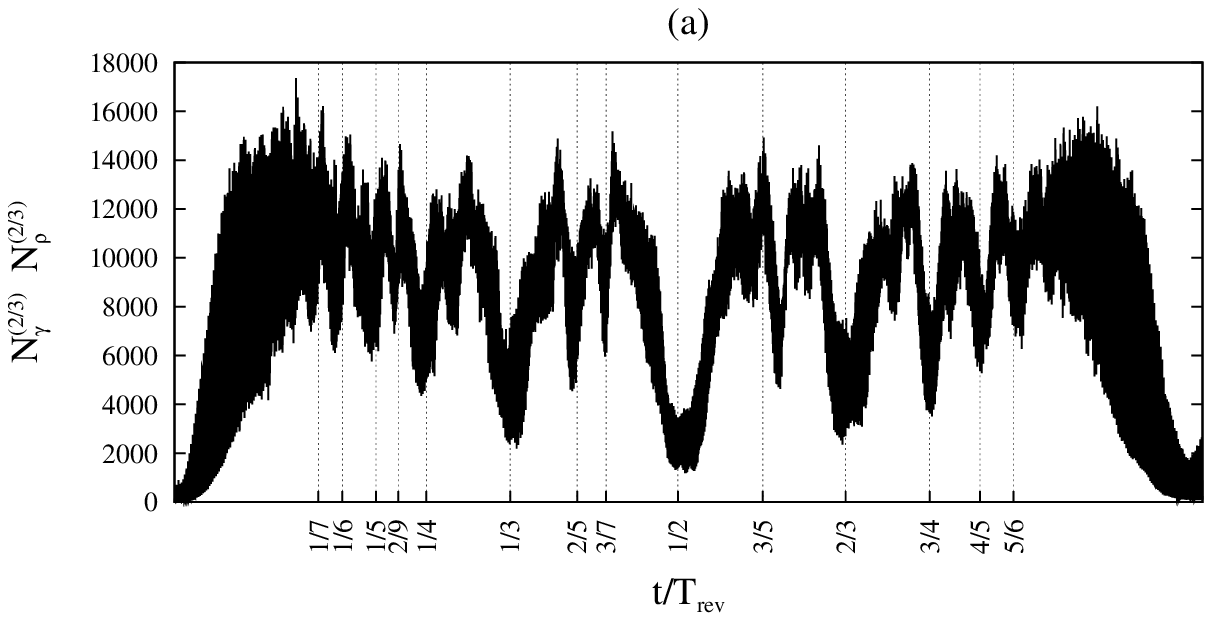}
\includegraphics[angle=0,width=.5\textwidth]{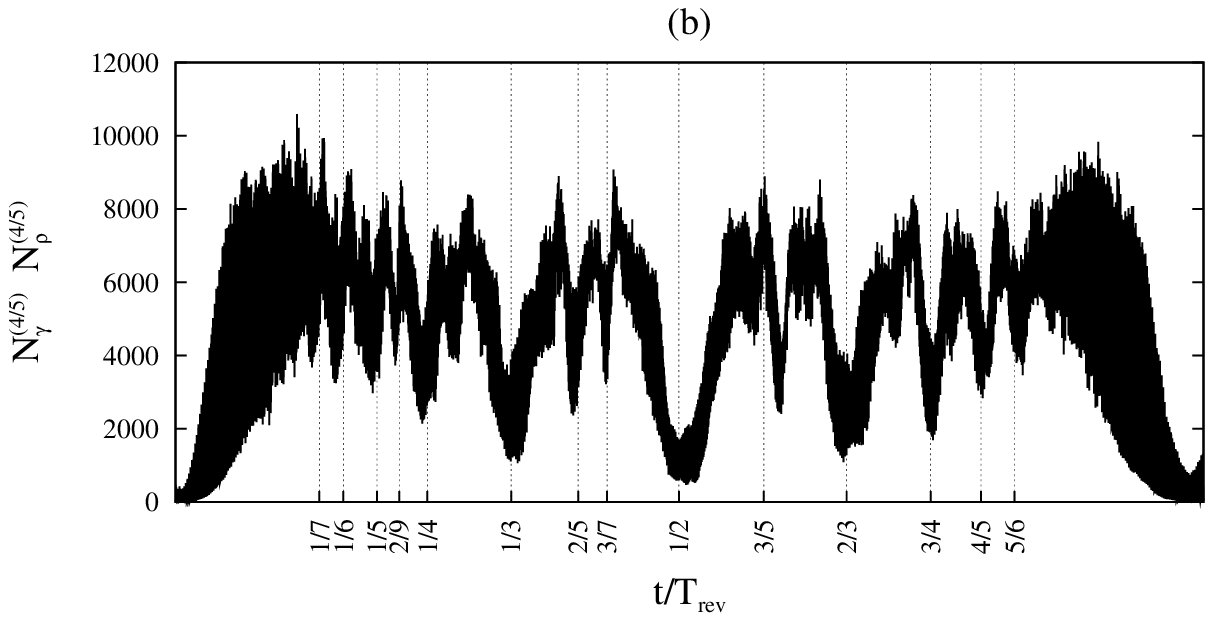}
\caption{Time dependence of $N^{(\alpha)}_\rho N^{(\alpha)}_\gamma$ and main fractional
revivals for a quantum bouncer. Parameters as in Fig. \ref{fig1}} 
    \label{fig3} 
  \end{center} 
\end{figure} 

\begin{figure}[tb] 
\begin{center} 
\includegraphics[angle=0,width=.5\textwidth]{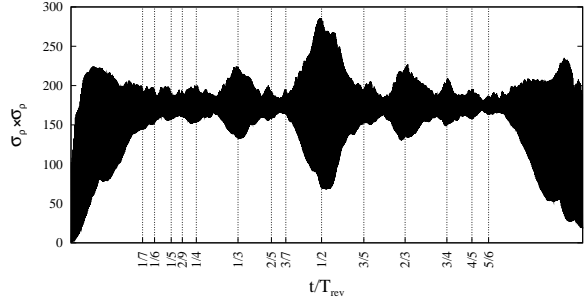}
\caption{Time dependence of $\sigma_\rho \sigma_\gamma$ and main fractional
revivals for a quantum bouncer. Parameters as in Fig. \ref{fig1}} 
    \label{fig4} 
  \end{center} 
\end{figure}

We have computed the temporal evolution of the uncertainty products \eq{uncertainty} 
and $\sigma_\rho \sigma_\gamma$ for the initial conditions $z_0=100$ and
$\sigma=1$. Figure \ref{fig1} displays $N^{(\alpha)}_\rho \sigma_\gamma^2$ and the location
of the main fractional revivals for  $\alpha=2/3$ (panel (a)) and  $\alpha=4/5$ (panel (b)).
Figures \ref{fig2} and \ref{fig3} show, respectively, $N^{(\alpha)}_\gamma \sigma_\rho^2$ and $N^{(\alpha)}_\rho N^{(\alpha)}_\gamma$
for the same values of $\alpha$ as in Fig. \ref{fig1}.
For comparison, we also show in Fig. \ref{fig4} the computed time evolution of $\sigma_\rho \sigma_\gamma$ for the same initial wave packet.
In every case it can be observed that the uncertainty products decrease and reach a minimum at most of the fractional revivals,
although the description provided by the products of entropies and of entropy and variance is more clear than that of the standard 
uncertainty product $\sigma_\rho \sigma_\gamma$. 

%
%

\section{Summary}

To summarize, we have studied the revivals and fractional revivals of a quantum bouncer by 
means of new uncertainty relations that combine time dependent dispersions and R\'enyi 
entropies. As it is also the case of other entropic approaches, it is found that they successfully account
for the wave packet regeneration. A comparison is made with the description provided by the standard variance-based
uncertainty product, to conclude that the entropic approach is generally superior. 

This work was supported in part by the
Spanish projects FIS2005-00973 (Ministerio de Ciencia y
Tecnolog\'ia), FQM-2725 and FQM-165 (Junta de Andaluc\'ia).

\end{document}